\documentclass[twocolumn]{aastex62}

\usepackage{ulem}
\usepackage{appendix}
\usepackage{amsmath}
\usepackage{amssymb}
\usepackage{threeparttable}

\defcitealias{xu22}{paper I}

\shorttitle{Planetesimal Formation in Dust Rings}
\shortauthors{Xu \& Bai}

\begin{document}


\title{Turbulent Dust-trapping Rings as Efficient Sites for Planetesimal Formation}

\author[0000-0002-2986-8466]{Ziyan Xu}\thanks{ziyan.xu@ens-lyon.fr}
\affiliation{Univ Lyon, Univ Lyon 1, ENS de Lyon, CNRS, Centre de Recherche Astrophysique de Lyon UMR5574, F-69230, Saint-Genis-Laval, France}
\affiliation{Kavli Institute for Astronomy and Astrophysics, Peking University, Yiheyuan 5, Haidian Qu, 100871 Beijing, China}
\affiliation{Department of Astronomy, Peking University, Yiheyuan 5, Haidian Qu, 100871 Beijing, China}

\author[0000-0001-6906-9549]{Xue-Ning Bai}\thanks{xbai@tsinghua.edu.cn}
\affiliation{Institute for Advanced Study, Tsinghua University, 100084 Beijing, China}
\affiliation{Department of Astronomy, Tsinghua University, 100084 Beijing, China}

\begin{abstract}

Recent observations of protoplanetary disks (PPDs) in the sub-mm have revealed the ubiquity of annular substructures, indicative of pebble-sized dust particles trapped in turbulent ring-like gas pressure bumps.
This major paradigm shift also challenges the leading theory of planetesimal formation from such pebbles by the streaming instability, which operates in a pressure gradient and can be suppressed by turbulence.
Here we conduct three-dimensional local shearing-box, non-ideal magnetohydrodynamic (MHD) simulations of dust trapping in enforced gas pressure bumps including dust backreaction.
Under the moderate level of turbulence generated by the magnetorotational instability (MRI) with ambipolar diffusion that is suitable for outer disk conditions, we achieve quasi-steady states of dust trapping balanced by turbulent diffusion.
We find strong dust clumping in all simulations near the gas pressure maxima, reaching a maximum density well above the threshold of triggering gravitational collapse to form planetesimals. A strong pressure bump concentrates dust particles towards bump center. With a weak pressure bump, dust can also concentrate in secondary filaments off the bump center due to dust backreaction, but strong clumping still occurs mainly in the primary ring around the bump center.
Our results reveal dust-trapping rings as robust locations for planetesimal formation in outer PPDs, while they may possess diverse observational properties.
\end{abstract}


\section{INTRODUCTION}

Planetesimals are $\sim1-1000$ km sized bodies in protoplanetary disks (PPDs) as building blocks of planets. 
Following the growth of dust grains to mm-cm sizes in PPDs, planetesimals are believed to form from such grown dust particles, representing the intermediate stage of planet formation. This stage is also considered to be the most difficult, as surface forces can no longer stick particles together upon collisions, nor does self-gravity which dominates among solids of larger sizes \citep{chiang10}. Generally, planetesimal formation is understood as a two-phase process, where particles first concentrate into high-density clumps, followed by their gravitational collapse. What drives the formation of particle clumps thus becomes the key to understanding planetesimal formation. Regardless of the specific mechanism, particles of mm-cm sizes strongly interact with gas via aerodynamic drag, and the gas dynamics in PPDs, especially disk turbulence, must play an important role.

The leading theory of planetesimal formation is the streaming instability \citep[SI,][]{youdin05}, resulting from the two-way drag force between gas and dust in the presence of a background pressure gradient in PPDs. Hybrid gas-dust simulations found that the non-linear evolution of the SI leads to dust clumping \citep[e.g.,][]{johansen09b,bai10b,carrera15,yang17}, provided the dust abundance $Z$ (ratio of dust to gas surface density) exceeds some threshold (typically $\gtrsim0.02$, but see \citealp{li21}). While highly successful, vast majority of SI simulations assumed no external turbulence. However, recent studies found that even modest level of external disk turbulence suppresses the SI \citep{umurhan20,chen20,gole20}, challenging the SI paradigm.

\begin{figure*}[t]
\begin{center}
\includegraphics[width=0.9\textwidth]{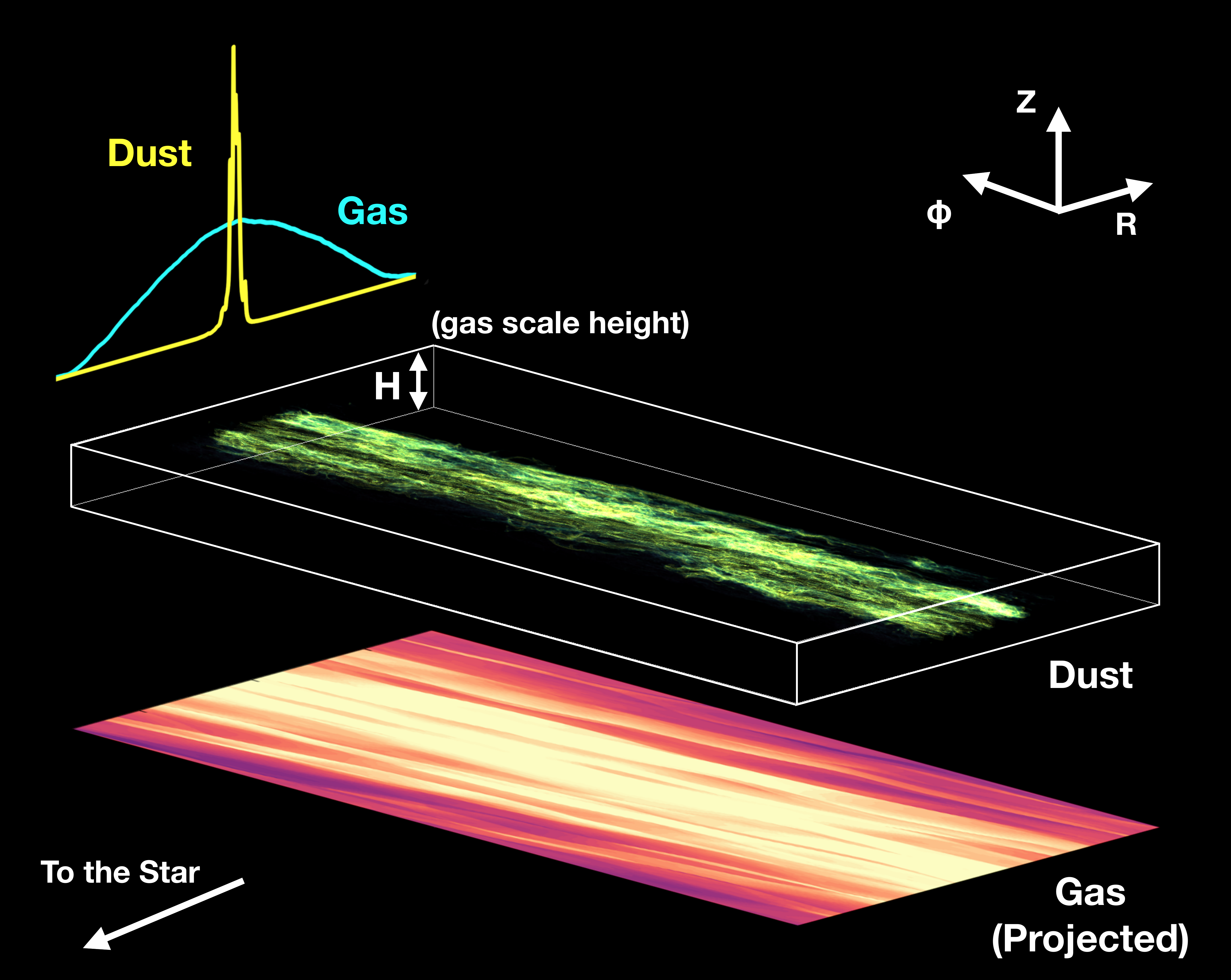}
\end{center}
\caption{
A snapshot of 3D visualization of the dust spatial distribution in our simulation with strong gas bump (run Z2H) at the time of $T=1380\Omega_K^{-1}$ after inserting dust particles. Dust density is mapped to brightness, where brighter region indicates higher dust density. Also shown in the figure are vertically averaged gas density projected to the $x$-$y$ ($R$-$\phi$) plane, and vertically and azimuthally averaged radial profiles of dust and gas density, with amplitudes of the profiles re-scaled for better illustration. \label{fig:intro}}
\end{figure*}

The outer disk regions ($\gtrsim20$ AU) are the most accessible to spatially-resolved disk observations. At sub-mm (where mm-sized dust radiates most efficiently), the prevalence of ring-like substructures has recently been firmly established \citep[e.g.,][]{andrews18,long18}, widely believed to be indicative of dust trapping in pressure bumps \citep{dullemond18}\footnote{Unless otherwise noted, ``pressure bump" refers to dust trapping pressure bumps in this paper, i.e., strong enough to overcome the background pressure gradient and generate local pressure maxima.}. Dust trapping in local pressure maximum is firstly demonstrated by \cite{whipple72}, and the formation mechanism of pressure bumps can be planet-disk interaction \citep[see review of][]{paardekooper22} or non-planet origins such as MHD zonal flows \citep[e.g.,][]{suriano18,riols20,cui21}. While there are direct and indirect evidence pointing to weak disk turbulence \citep{pinte16,flaherty17,flaherty18,teague18}, modeling dust ring widths suggest modest level of turbulence \citep{dullemond18,rosotti20}. Together, these observational constraints lead to the emerging picture that most mm-sized dust particles in outer PPDs are most likely concentrated in weakly-to-modestly turbulent, largely-axisymmetric gas pressure bumps. The strongly enhanced local dust abundance in such pressure bumps yields favorable condition for planetesimal formation. However, given that the SI is unexpected to operate without pressure gradient, and can be suppressed by external turbulence, it is unclear whether it will further lead to dust clumping and hence planetesimal formation.

To address the feasibility of planetesimal formation in more realistic environment of turbulent dust rings, we conduct hybrid gas-particle simulations in the standard shearing-box framework that mimics a local patch of the disk. It is anticipated that the magneto-rotational instability (MRI) is one primary mechanism for driving turbulence in the outer PPDs. Due to very low level of disk ionization, the MRI is weakened by ambipolar diffusion (AD) as the dominant non-ideal magnetohydrodynamic (MHD) effect in the outer disk. Turbulence in our simulations are self-consistently generated by the MRI with AD. We further implemented a novel forcing prescription to drive a pressure bump in the center of our simulation box with essentially no impact on disk turbulence.
Our simulations generalize from our earlier work \citep[][hereafter \citetalias{xu22}]{xu22} by imposing and enforcing the pressure bump, allowing us to achieve a steady-state configuration with dust trapping towards pressure bumps balanced by dust diffusion in realistic disk turbulence in a self-consistent and well-controlled manner. They differ from the recent works of \citep[][]{carrera21,carrera22a,carrera22b}, who conducted simulations with a similar procedure to enforce the pressure bump but without external turbulence, thus cannot achieve such steady-state balance of dust trapping and diffusion as envisioned in recent disk observations \citep{dullemond18,rosotti20}.

This letter is organized as follows. 
We first introduce the implementation of the pressure bump and simulation setup in \S \ref{sec:method}, then present our main results in \S \ref{sec:results}. In \S \ref{sec:discu}, we will further analyze the results, focusing on dust clumping conditions and ring dynamics, followed by comparison with recent studies and discussion on observational implications. We summarize and discuss future prospects in \S \ref{sec:conclusions}. 

\begin{table}[t]
    \centering
     \caption{List of simulation runs.}
    \label{tab:sumsetup}
    \begin{tabular}{lcccccccc}
   Runs  & $Z$ & $N_x \times N_y \times N_z$  & $A_{bump}$ & $H_d/H$ & $w_{d}/H$ \\
   \hline
   Z0 & 1e-30 & 512 $\times$ 512 $\times$ 64  & 0.5 & 0.067 & 0.55\\
   Z2  & 0.02 & 512 $\times$ 512 $\times$ 64  & 0.5 & 0.014 &0.16\\
   Z0H & 1e-30 & 1024 $\times$ 1024 $\times$ 128  & 0.5 & -&-\\
   Z1H  & 0.01 & 1024 $\times$ 1024 $\times$ 128  & 0.5 & 0.025 &0.25\\
   Z2H  & 0.02 & 1024 $\times$ 1024 $\times$ 128 & 0.5 & 0.017 &0.16\\
      \hline
   Z0w & 1e-30 & 512 $\times$ 512 $\times$ 64 & 0.25 & 0.075 &0.83\\
   Z1w  & 0.01 & 512 $\times$ 512 $\times$ 64 & 0.25 & 0.031 &0.87\\
   Z2w  & 0.02 & 512 $\times$ 512 $\times$ 64 & 0.25 & 0.022 &0.74\\
   \hline
    \end{tabular}
    \begin{tablenotes}
    \item All simulations have a boxsize of $8H \times 16H \times H$, with $\beta_0=12800$, $Am=2$. Gas bump width  $w_{bump} = 2H$, with no background pressure gradient. Particle number is 2 per cell on average, with particle stopping time $\tau_s=0.1$.
    \item All parameters in this table are initial setups, except that $H_d$ and $w_{d}$ are measured from simulation. 
    \end{tablenotes}
\end{table}

\section{Method and Simulations} \label{sec:method}

We use the Athena MHD code \citep{stone08} to perform 3D hybrid gas-particle local shearing-box simulations considering non-ideal MHD with ambipolar diffusion. 
The formulation and basic simulation setup is similar to \citetalias{xu22}, which we only describe briefly here. The $x, y, z$ coordinates in a local shearing-box correspond to radial, azimuthal and vertical directions in a global disk, as illustrated in Figure \ref{fig:intro}. We use an isothermal equation of state with sound speed $c_s=1$ and local Keplerian frequency $\Omega_K=1$ in code units, which define our length unit of disk scale height $H=c_s/\Omega_K=1$. Gas density, velocity and pressure are denoted by $\rho, {\boldsymbol v}$ and
$P=\rho c_s^2$, respectively.
Dust particles are characterized by their dimensionless stopping time $\tau_s\equiv t_s\Omega_0$, and the height-integrate dust-to-gas mass ratio $Z$ for background gas surface density. Our simulations focus on the midplane region, and hence we neglect vertical gravity in the gas (unstratified), but vertical gravity is included for particles to allow dust settling. We impose a net vertical magnetic flux characterized by $\beta_0=12800$, the ratio of background gas pressure to magnetic pressure of the net vertical field. Ambipolar diffusion is characterized by the Ambipolar Elsasser number $Am$, where we fix $Am=2$ appropriate for typical outer disk conditions.

\subsection{Implementation of the gas pressure bump} \label{subsec:formulation}

On top of the simulation setup in \citetalias{xu22}, we further implement a gas pressure bump by a prescribed forcing procedure.
Let $\rho_0=1$ be the constant background gas density. The forcing will redistribute the density, and we aim to achieve a Gaussian-like density (and hence pressure) bump in our simulation box of the following form
\begin{equation}
\frac{\rho(x)}{\rho_0}= 1+ A_{\rm bump}\exp\bigg[-\frac{x^2}{2w_{\rm bump}^2}-0.2\bigg(\frac{x^2}{2w_{\rm bump}^2}\bigg)^4\bigg],\\
\end{equation}
where
$A_{\rm bump}$ is the dimensionless amplitude of the pressure bump, $w_{\rm bump}$ is the effective width of the bump. The first term in the exponential gives the overall Gaussian profile, while the second term damps the profile towards the domain edges to satisfy the shearing-periodic radial boundary condition.

The radial pressure gradient associated with this bump must be balanced by an azimuthal flow:
\begin{equation}
    \frac{v'_{y,{\rm bump}}(x)}{c_s} = 
    \frac{H}{2}\frac{d\ln\rho(x)}{dx}\ ,
\end{equation}
where the prime $'$ in $v_y$ denotes deviation from the Keplerian velocity $v_K=-(3/2)\Omega_Kx$.
Note that in our simulations, as we focus on regions near the pressure maxima, we do not impose any background pressure gradient.

To generate and sustain the bump,
we apply an azimuthal forcing in our simulation, so that the azimuthal velocity relax to the desired profiles within a timescale of $t_{\rm relax}=60\Omega_K^{-1}$:
\begin{equation}
\Delta v'_y(x) = [v_{y,{\rm bump}}(x) - \overline{v'_y}(x)] [1-\exp(-\Delta t/t_{\rm relax})],
\end{equation}
where $\overline{v'_y}(x)$ is azimuthally and vertically averaged. Our forcing procedure is analogous to that of \cite{carrera21}, but with several differences. First, our forcing acts on azimuthal velocity (momentum) only, thus is mass conserving. It directly mimics the torque exerted on the disk responsible for making the bump
\citep[be it of planet origin, or non-planet origin as in][]{cui21},
and the desired density profile is a direct consequence from this forcing. Second, our forcing is homogeneously applied in the $y$ and $z$ directions, as opposed to relaxation {\it at individual grid cells} which introduces local friction. Third, our relaxation time is much longer, as opposed to $\sim1\Omega_K^{-1}$. These facts minimizes the impact of our forcing procedure to disk turbulence, especially avoiding artificial damping at small scales.\footnote{Our choice of relaxation time is generally consistent with the typical timescale for magnetic flux evolution (and thus bump evolution) indicated in \cite{cui21}. We further tested $t_{\rm relax}=30\Omega_K^{-1}$ and $120\Omega_K^{-1}$ respectively (not shown in paper), and the result is not sensitive to the change of $t_{\rm relax}$ in our simulation in this paper. }

\subsection{Simulation Procedures and Parameters} \label{subsec:simsetup}

The list of our simulation runs are shown in Table \ref{tab:sumsetup}.
We explore strengths of the gas pressure bump, presence/absence of dust feedback, different levels of solid abundance, and simulation resolutions. The simulation setup and parameters are described as follows.

We first run simulations without dust particles for a time of $t_0 = 120 \Omega_K^{-1}$ for the MRI to grow and fully saturate into turbulence, following our previous work (\citetalias{xu22}).
The gas pressure bump is initiated at the beginning of the simulation, with bump width $w_{\rm bump} = 2H$. 
We explore two cases: strong and weak bumps, with bump amplitude $A_{\rm bump} = 0.5$ and $0.25$ separately, and the  weak bump runs are marked by ``w'' in the run name.\footnote{We have examined that these bump amplitudes are below the threshold for triggering the Rossby wave instability (RWI) for all azimuthal wave numbers under global settings \citep{ono16}, and we anticipate the presence of the MRI turbulence also makes our pressure bump less prone to the RWI.} Our choice of bump width generally agrees with observational constraints \citep{dullemond18,rosotti20}, and both the bump width and amplitude in our simulation are reasonably consistent with the results of \cite{cui21}.

\begin{figure*}[t]
\begin{center}
\includegraphics[width=0.9\textwidth]{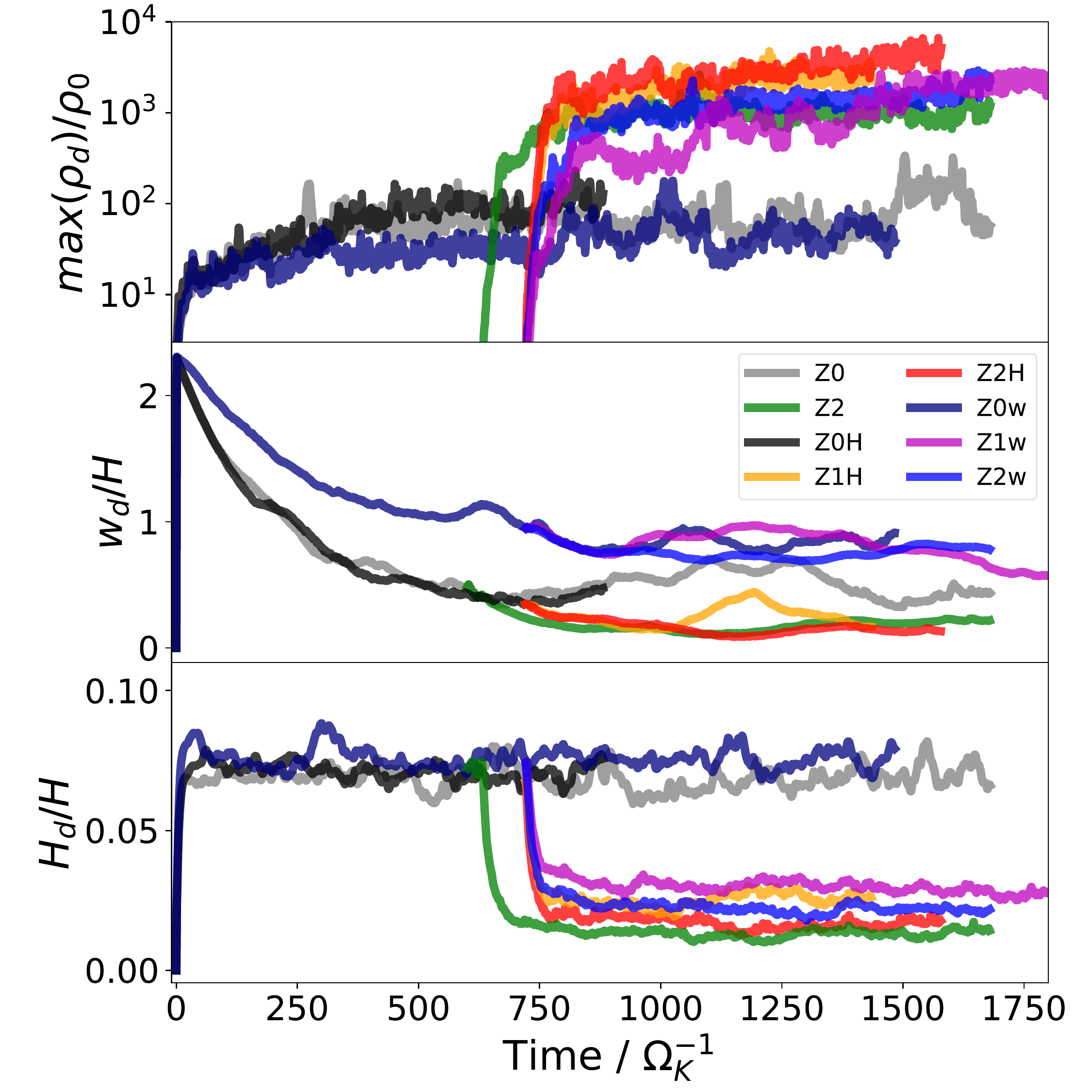}
\end{center}
\caption{
Time evolution of dust maximum density (upper panel), dust ring width (middle panel), and dust scale height (lower panel) for all our simulation runs. Dust maximum density for no-feedback runs (Z0, Z0H, and Z0w) are scaled to $Z=0.02$. \label{fig:hist}}
\end{figure*}

Dust particles are then inserted into the simulation box at the time of $t_0 = 120 \Omega_K^{-1}$ without backreaction to the gas, with random horizontal positions and a vertical distribution following a Gaussian profile $\propto \exp(-z^2/2H_{\rm d0}^2)$, with $H_{\rm d0} = 0.02 H$. We define $T=T_{\rm sim}-t_0$, where $T_{\rm sim}$ is the actual simulation time, and use $T$ instead of $T_{\rm sim}$ throughout this paper.
We use a single particle species with fixed stopping time $\tau_s=0.1$, 
corresponding to mm-cm size particles in the midplane regions of the outer disk in standard disk models \citep[e.g.,][]{chiang10}.
As dust particles passively follow the gas,
settling balanced by turbulent diffusion in the vertical direction makes them maintain certain scale height (see \citetalias{xu22}), while the balance between trapping toward pressure maxima and turbulent diffusion in the radial direction allow them to concentrate into a ring with certain ring width $w_d$ (see Figure \ref{fig:hist}), approaching a steady state after $T=600\Omega_K^{-1}$.
Next, we gradually introduce dust feedback to the gas by increasing dust abundance gradually to the desired value with over a time interval of $30 \Omega_K^{-1}$.
We consider two {\it global} dust abundances $Z=0.01$ and $Z=0.02$,
and these runs are labeled by ``$Z$" followed by a number indicating the $Z$ values.
In all simulations, there are on average 2 particles per cell, ensuring sufficient particle statistics for studying dust dynamics. 

We choose simulation box size to be $8 H\times 16 H \times H$ in $x$, $y$ and $z$ directions respectively.
We choose a fiducial resolution of 64/H (in the $x$ and $z$ directions; the resolution in the $y$ direction is half of that).
We further conducted simulations with a higher resolution of 128/H for comparison, labeled with an ``$H$" at the end of the run names.

\begin{figure*}[t]
\begin{center}
\includegraphics[width=.9\textwidth]{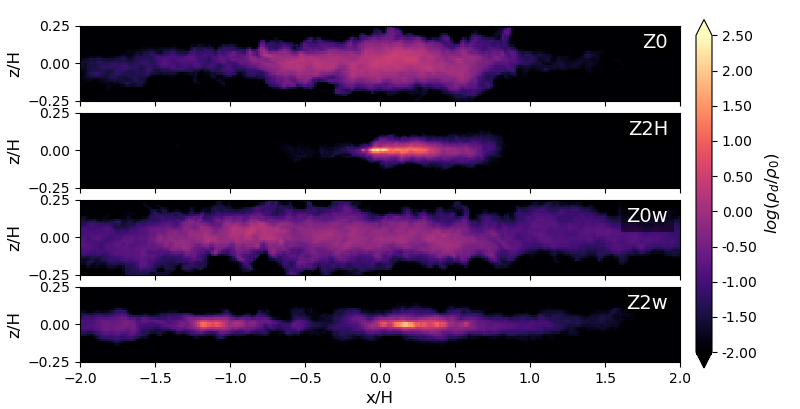}
\includegraphics[width=.9\textwidth]{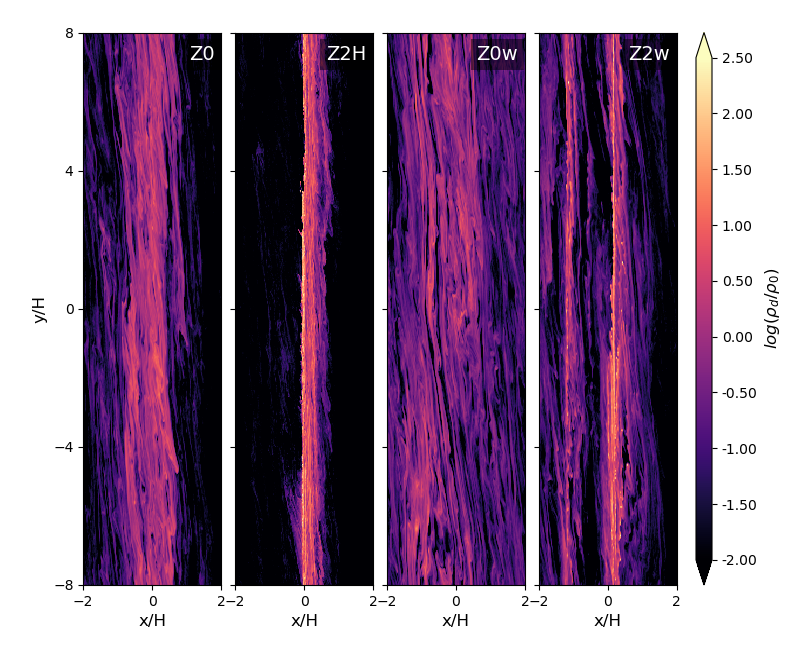}
\end{center}
\caption{Snapshots of projected dust density in the $x$-$z$ (upper four panels) and $x$-$y$ (lower four panels) planes for runs Z0, Z2H, Z0w, Z2w, at time $T=1380\Omega_K^{-1}$, when the secondary filament in Z2w is clearly seen. Dust density in the $x$-$z$ plane is azimuthally averaged over the whole domain, while in the $x$-$y$ plane, dust density is vertically averaged around the midplane ($z = \pm 0.05H$) to highlight dust structures.  \label{fig:xyxz}}
\end{figure*}

\section{Results} \label{sec:results}

An overview of our simulations is shown in Figure \ref{fig:intro}, featuring a 3D visualization of the dust spatial distribution for our run Z2H at time $T=1380\Omega_K^{-1}$.  
Our forcing procedure generates a wide and steady gas bump, with additional density fluctuations (i.e., density waves) as a result of the MRI turbulence \citep{heinemann09}. Clearly, the dust concentrates to the center of the gas bump, whose width is much narrower than the gap bump width. Besides settling and trapping, and the dust distribution is highly non-smooth and clumpy under background turbulence.

In this section, we present our main simulation results.
We measure the maximum dust density $\rho_{d, {\rm max}}$, normalized by background gas density $\rho_0$, as a proxy for dust clumping. Dust density in no-feedback runs (Z0 and Z0H) are rescaled to Z=0.02 for better comparison. For reference, we consider $\rho_{d, {\rm max}}\gtrsim200\rho_0$ as a threshold condition for triggering gravitational collapse, and hence planetesimal formation, which is applicable for typical outer disk conditions (see Equation 16 and Section 5.1 in \citetalias{xu22}). We further measure the dust ring width $w_d$ and dust scale height $H_d$, defined as the rms value of the radial and vertical coordinates for all dust particles. The time history of $\rho_{d, {\rm max}}/\rho_0$, $w_d$ and $H_d$ from all simulations is shown in Figure \ref{fig:hist}. The time-averaged values of $w_d$ and $H_d$ are presented in Table 1, averaged within $T=800-1500\Omega_K^{-1}$. These are accompanied by Figure \ref{fig:xyxz}, where we show the projected dust density in the $x-y$ and $x-z$ planes for Z0, Z2H, Z02 and Z2w runs at the saturated states.
We find that the results for the strong bump and weak bump cases are distinct, which we discuss separately below.

\subsection[]{The strong bump case: dust clumping in a single thin ring}\label{subsec:strong}

Top panel of Figure \ref{fig:hist} shows that dust density presents modest concentration in the absence of dust feedback, with $\rho_{d, {\rm max}}/\rho_0$ reaching $\sim100$, approaching yet barely reach the threshold for triggering planetesimal formation. This is to due to a combination dust trapping and intrinsic fluctuations. Reading from the second and third panels of Figure \ref{fig:hist}, we find a dust ring width of $w_d\sim0.55H$, and a scale height of $H_d\sim0.067H$ (comparable to $\sim 0.08H$ found in \citetalias{xu22} without enforced pressure bump). This yields a midplane density enhancement factor of $\sim50$ at bump center. Indeed, as shown in Figure \ref{fig:xyxz}, the mean midplane density at bump center for run Z0 (renormalized assuming $Z=0.02$) is about $\overline{\rho_d}\sim1$. We do not observe obvious dust clumping from the averaged dust densities along the vertical and azimuthal directions, and thus attribute an additional factor of $\sim50-100$ enhancement in $\rho_{d, {\rm max}}$ from localized density fluctuations. The results are largely independent of resolution, where $\rho_{d, {\rm max}}$, $w_d$ and $H_d$ are consistent with each other between run Z0 and Z0H within fluctuation levels.

Turning on dust feedback leads to a rapid boost in $\rho_{d, {\rm max}}/\rho_0$, well exceeding $10^3$ (and hence the Roche density) for all runs with Z=0.01 (Z1H) and Z=0.02 (Z2 and Z2H), which is therefore expected to trigger planetesimal formation. There are also visible bright spots seen in the bottom panel of Figure \ref{fig:xyxz} for run Z2H, suggesting that such clumping is physical instead of random Poisson fluctuations. There is a trend of higher resolution leading to higher level of dust clumping, where for runs Z1H and Z2H, $\rho_{d, {\rm max}}/\rho_0\gtrsim5000$, a factor of $\sim50$ enhancement compared to the no feedback case. We have further examined the cumulative probability density distribution (CDF, see Appendix \ref{appsec:cdf}), which reveal that dust density is indeed substantially boosted with a significant fraction (a few tens of percent) residing in over-Roche-density regions.

Somewhat surprisingly, with dust feedback, we observe a dramatic reduction of the ring width to $w_d\approx0.2H$,
$\sim 3$ times narrower than that without feedback. Moreover, the thickness of the dust layer is also reduced substantially by a factor of $3-4$. Examining Figure \ref{fig:xyxz} reveal that this reduction is not homogeneous: there is a strong/spiky dust concentration and more substantial settling at the bump center (see also the top left panel of Figure \ref{fig:vy_x1} and Figure 3 in \citetalias{xu22}), which reduces the measured $w_d$ and $H_d$ towards lower values. We note that the reduction of $H_d$ was also reported in \citetalias{xu22}, which was mainly attributed to reduction of turbulence correlation time.
However, in the presence of pressure bump, $H_d$ is further reduced by a factor of nearly $\sim 2$. 
The dramatic reduction of ring width, as well as further promotion of dust settling, are likely the outcome of strong dust mass loading in the bump center (see further discussions in \S \ref{subsec:discu.interplay} and Appendix \ref{appsec:diff}).
The results are convergent with resolution, as the values of $w_d$ and $H_d$ are consistent with each other between runs Z2 and Z2H.

\subsection[]{The weak bump case: diverse ring properties}

The most prominent feature for the weak bump case compared to the strong bump case is the dust ring properties.
The rest of the results, especially dust clumping and settling, are similar to the case with a strong bump.

The middple panel of Figure \ref{fig:hist} shows that without dust feedback, $w_d$ in the weak bump case is generally higher compared to the strong bump case by a factor of up to $\sim2$. This is easily accounted for due to the reduced gas pressure gradient in the two sides of the gas bump, resulting in weaker dust trapping. This is also seen in Figure \ref{fig:xyxz}, showing a smooth and radially extended distribution of dust from run Z0w, together with azimuthally extended stripes akin to the MRI density waves.

When turning on dust feedback, interestingly, the dust ring width $w_d$, measured from the rms radial coordinates, remains similar for all cases.
However, when examining the dust spatial distribution in Figure \ref{fig:xyxz}, we find that
instead of concentrating into a single ring at the bump center, dust dynamics in weak bump is more complicated, potentially forming multiple fine-scale filaments within the gas bump. In the case of run Z2w, we observe two dust rings where the primary is close to bump center, and a secondary ring more than $H$ away from the bump center, in addition to a few minor filaments.
The formation of this more complex ring/filamentary structure is likely caused by the interplay between dust feedback and gas dynamics within the bump. In addition, these structures generally stay while slowly evolve over tens to hundreds of $\Omega_K^{-1}$. Details of the formation and evolution of (multiple) fine-scaled ring structures will be further discussed in \S \ref{subsec:discu.interplay} and Appendix \ref{appsec:longterm}.

\section{Discussion} \label{sec:discu}

\subsection{Clumping conditions}
\label{subsec:discu.condition}

\begin{figure*}[t]
\begin{center}
\includegraphics[width=.9\textwidth]{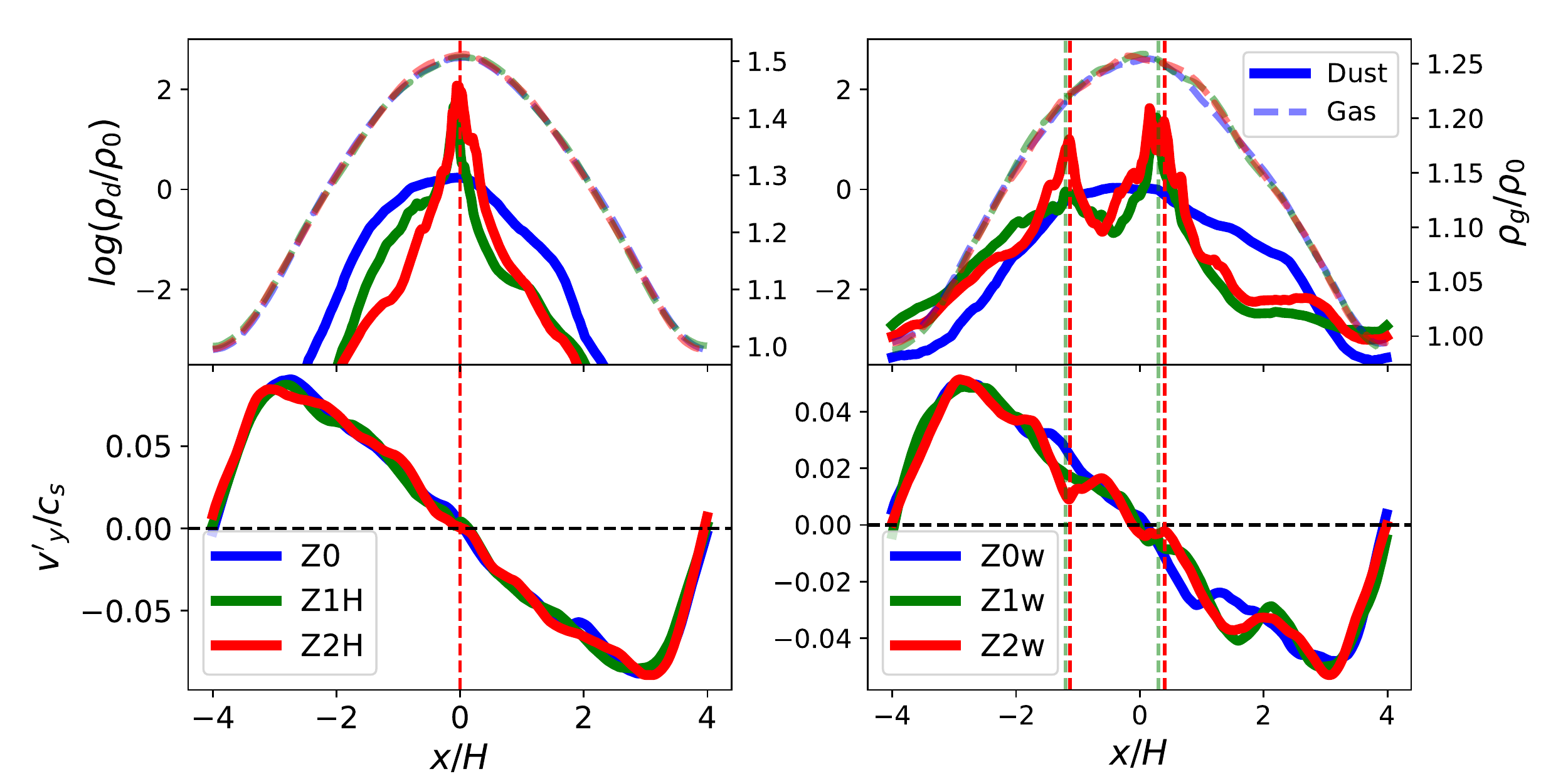}

\includegraphics[width=.9\textwidth]{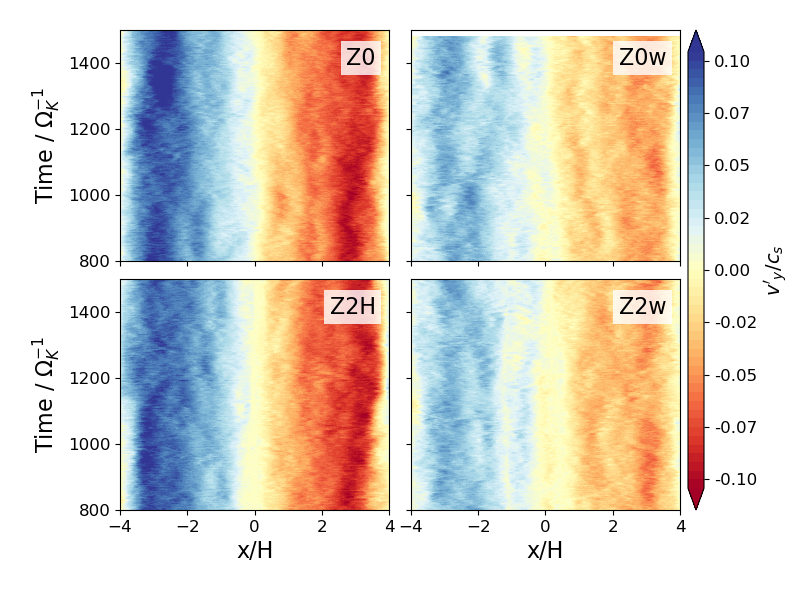}
\end{center}
\caption{Upper four panels: radial profiles of vertically and azimuthally averaged dust density ($\rho_d$, solid curves in upper two panels), gas density ($\rho_g$, dashed curves in upper two panels), and Keplerian-subtracted azimuthal gas velocity ($v'_y$, solid curved in lower two panels), at the midplane ($\pm 1$ cell from $z=0$), averaged between $T=800-1500\Omega_K^{-1}$. Vertical dashed lines mark the locations of the dust rings (peak of dust density), with colors corresponding to that used for radial profiles of each run.\label{fig:vy_x1} \\ Lower four panels: time evolution of the radial profile of $v'_y$, between $T=800-1500\Omega_K^{-1}$.}
\end{figure*}

For both the strong and weak bumps, sufficient dust clumping is seen to trigger planetesimal formation. Here we further discuss the clumping conditions to gain insight into the clumping mechanism. In doing so, we show in the top four panels of Figure \ref{fig:vy_x1} the radial profiles of vertically and azimuthally-averaged dust density, gas density and Keplerian-subtracted azimuthal gas velocity $v'_y$, at the very midplane of our simulation ($\pm$ 1 cell). The latter two are useful for quantifying the pressure gradient, and we further show the time evolution of the radial profiles of $\overline{v'_y}$ in four representative runs in the bottom panels.

\paragraph{Dust feedback}

Dust feedback is required for dust clumping in our simulations, as no clumping is seen in our run without feedback (Z0 and Z0w). 
As seen in the upper panels of Figure \ref{fig:vy_x1}, the dust midplane density in the no-feedback cases generally exhibits a Gaussian-like radial profile, conforming to the standard expectation of concentration balancing turbulent diffusion \citep{dullemond18}. Including dust feedback, the radial density profiles substantially deviate from Gaussian, showing density cusps at the location of dust rings, where dust density is enhanced by a factor of near $100$ relative to the no-feedback counterparts. The cusps are associated with the locations of dust clumping, and can be the consequences of dust clumping itself.

\paragraph{Solid abundance}
Dust clumping occurs in all simulations for mean solid abundance $Z$ (averaged over the simulation domain) as low as $0.01$. Note that due to dust trapping, the effective solid abundance (i.e., vertically integrated dust-to-gas mass ratio) in the bump region is much higher.
We may estimate the enhancement factor to be $\sim 4-8$ in the weak bump and strong bump cases based on the value of $w_d$, corresponding to $Z\gtrsim0.04$ around bump center in all our simulations.
Therefore, the overall solid abundance $Z$ in the bulk disk appears less relevant, as dust trapping can easily bring sufficient amount of dust to the bump region.

\paragraph{Local pressure maxima}

In the strong bump case (runs Z1H, Z2H), dust clumping primarily occurs at the central cusp, which is marked in a dashed line in Figure \ref{fig:vy_x1}. We further see from the radial profiles of $\rho_g$ and $v'_y$ that this location exactly coincides with the location of the gas pressure maxima, corresponding to $v'_y=0$.
In the weak bump case, the primary ring close to the bump center is also located
at the pressure maxima, which in this case is widened into a plateau with a width of $\sim0.5H$ as seen from the gas density profile as well as an extended region with $v'_y\approx0$. The cusp where strong dust clumping occurs is located within the plateau, again consistent with the requirement of local pressure maxima, or at least a zero-pressure-gradient region being a necessary condition for strong dust clumping.
These findings are consistent with the results reported in \citetalias{xu22}, where the pressure bumps were not enforced but generated by the MRI zonal flows (albeit somewhat unrealistic in local unstratified simulations).

\paragraph{Weak clumping in the secondary ring}
For the weak bump case of Z1w and Z2w, in addition to the primary ring near the bump center, an off-centered ``secondary'' ring is identified at $x \sim -1.2H$. 
We show in Appendix \ref{appsec:cdf} that the dust density CDF in the primary ring is similar to that in the strong bump case, with strong clumping. However, the dust CDF of the secondary ring is significantly different. The mean dust density is more than a factor of $\sim10$ lower, and only a very small fraction ($\lesssim3\%$) of particles reside in over-Roche-density regions (which exhibit as individual spots seen in the bottom right panel of Figure \ref{fig:xyxz}), indicating that clumping is much less efficient.
Examining the density (and hence pressure) profiles in the top right panel of Figure \ref{fig:vy_x1}, the location of the secondary ring is clearly not at the pressure maximum.

Overall, for both strong and weak bump cases, dust clumping is tightly related to the presence of pressure maxima as long as dust backreaction is included. While secondary rings/filaments can form for the weak bump case, they generally reside outside of the pressure maxima and do not show prominent/efficient clumping.

\subsection{Interplay between dust feedback and gas bump}
\label{subsec:discu.interplay}

\paragraph{Bump/ring profile}

In addition to dust clumping, the interaction between dust and gas also influences gas bump properties, leading to diversity in dust structures, as we discuss here.

Dust feedback changes the radial profile of the gas bump.
This can be seen in the yellow region in the time-evolution map of $v'_y$ at the midplane in Figure \ref{fig:vy_x1} by comparing the upper and lower panels. In the strong bump case, as well as around the primary ring in the weak bump case, dust feedback leads to a plateau of Keplerian rotation at the bump center, in comparison to that in run Z0 and Z0w. This is consistent with the previous finding in \citep{huang20}.

The secondary dust ring, despite not in the pressure maxima, can be sustained for at least dozens of orbits without drifting towards the bump center.
Examining the profile of $v'_y$, we find that its value shows a dip at the secondary dust filament, implying near-Keplerian rotation. This is the result of feedback from the dusty filaments modifying the rotational profile of the background gas. It occurs in our weak bump runs likely because the gas pressure gradient is not strong enough to force the heavily loaded dust filaments to drift to bump center, resulting in dust trapping.\footnote{
One might argue that the formation of secondary dust rings seen in run Z1w and Z2w be a consequence of the special timing when we turn on dust feedback due to pre-existing local density enhancements from the MRI turbulence (e.g., zonal flows). We thus further perform a simulation where dust feedback is turned on at a much later time $T_{sim}=1200\Omega_K^{-1}$.
We still find similar secondary dust rings/filaments in this case.}
Such a location shares similarities with a pressure maxima in that there is almost no radial or azimuthal drift between gas and dust, and is likely responsible for the weak clumping discussed in \S \ref{subsec:discu.condition}.
These secondary rings also evolve over the timescale of tens to hundreds of orbits,
as we discuss further in Appendix \ref{appsec:longterm}. 

\paragraph{Turbulent diffusion}
The enhanced dust mass loading in turbulent dust rings also affects the properties of background MRI turbulence.
With feedback included, we find that the gas rms velocities in both the radial and vertical directions are largely reduced in the dust ring (see details in Appendix \ref{appsec:diff}).
Such reductions can sufficiently explain the results reported in Section \ref{subsec:strong}, where the dust rings in the strong bump case becomes much narrower when dust feedback is included, and that dust becomes even more settled than that reported in \citetalias{xu22}.

\subsection{Comparison with other recent studies}

Recently, \cite{carrera21,carrera22a,carrera22b} conducted a series of hydrodynamic simulations of the SI, imposing a pressure bump by a similar (albeit more dissipative) forcing procedure as ours but without external turbulence. However, the pressure bump in most of their simulations are not sufficiently strong to overcome the background pressure gradient. As a result, particles get accumulated (but {\it not trapped}) in regions with lower pressure gradient, making such regions the most favorable for clumping and planetesimal formation \citep[e.g.,][]{bai10c}. In other words, their simulations identified the favorable conditions for dust clumping by the conventional SI in the presence of radial-dependent pressure gradient. Our simulations, on the other hand, focus on planetesimal formation in {\it dust-trapping rings}, i.e., the pressure maxima. The problem is only well-defined in the presence of external turbulence, as expected from observations \citep{dullemond18,rosotti20}, otherwise, dust would indefinitely get concentrated towards ring center over the trapping timescale (this issue may also apply to the analytic work of \cite{auffinger18}), forming planetesimals along the way. Overall, the physical picture in our simulations is distinctly different from the works of \cite{carrera21,carrera22a}, and likely better reflect outer disk conditions. Together with \citetalias{xu22}, we showed that for the MRI turbulence, the condition for planetesimal formation requires the presence of pressure maxima, again different from the conventional SI scenario.

The outer disk may also be subject to turbulence driven by the vertical shear instability \citep[VSI,][]{nelson13}. Using 2D global simulations, \cite{schafer20} found that the SI can co-exist with the VSI, and the VSI even facilitates the development of the SI and dust clumping, likely due to the overdensities induced by the VSI. We note that the overdensities from the VSI itself is also insufficient to overcome the background pressure gradient.
This is different from the case with the MRI, where we found that dust clumping requires the presence of pressure maxima, suggesting that dust clumping conditions likely depends on the nature turbulence.
\cite{lehmann22} found that the presence of an initial pressure bump (but without forcing) further facilitates dust concentration, though their resolution is insufficient to capture dust clumping. Moreover, the VSI leads to vortex formation in 3D \citep{manger20}, and hence dust clumping in dust-laden vortices \citep[e.g.,][]{raettig21}, which represents another alternative for planetesimal formation in turbulent disks. It is worth noting that the VSI and the MRI can co-exist (Cui \& Bai 2022, submitted), and turbulence is dominated by the MRI/VSI when the ambipolar Elsasser number $Am$ is large/small compared to unity. It is likely that clumping conditions still depend on the nature of turbulence (be it due to the VSI or the MRI), although more investigations are needed to understand the clumping mechanisms, and how the dust dynamics is affected when both instabilities are present.

\subsection{Observational implications}

\paragraph{Dust clumping and fine-scale substructures}

There can be multiple observational diagnostics to probe the turbulent dust rings. First, as such dust rings are clumpy, there can be both optically thick and optically thin regions within the ring. Observations at sub-mm by ALMA can hardly resolve such regions, but the clumpiness could be reflected in the spectral indices, as suggested by \cite{scardoni21}. Further analysis of our results with radiative transfer and comparison with ALMA observations may provide evidence of dust clumping in the observed dust rings.

Second, we have seen the formation of secondary rings/filaments in our weak-bump simulations. These fine-scale substructures are separated on scales $\lesssim H$, but are all located within a single gas pressure bump. We note that generally we do not expect ring-like substructures in the gas component on scales $\lesssim H$ \citep{dullemond18}, whereas observationally, evidence of such fine-scale ring-like substructures is mounting \cite{jennings22}, with hints of multiple rings within previously identified rings. Our results provide a natural explanation of forming multiple sub-$H$ scaled substructures within a single gas bump. For instance, they may provide an alternative explanation to the compact double rings observed in HD 169142 at 57 and 64 AU separately, previously interpreted as the consequence of a migrating protoplanet \citep{perez19}.

Third, we note that the dust rings from our simulations are not entirely axisymmetric, with fine-scale substructures along the azimuth. Although our simulations are local and do not cover the entire azimuth, we notice that fine-scale, low-contrast asymmetries have been observed in some systems \citep[e.g., DM Tau,][]{hashimoto21} and may be common among gapped protoplanetary disks \citep{vandermarel21}.

\paragraph{Disk turbulence}

Dust ring width has recently been used to constrain the level of turbulence (the $\alpha$ parameter) in disks \citep{dullemond18,rosotti20}.  
However, as discussed in \S \ref{subsec:discu.interplay}, disk turbulence in the presence of a dust ring can be both anisotropic and non-uniform due to dust feedback.

More specifically, turbulent diffusion in the radial direction in our simulations is overall stronger than that in the vertical direction, leading to a large aspect ratio of $w_d/H_d \sim 10$ for the single dust ring in run Z2H, which is 5 times of aspect ratio of the gas bump inserted in the simulation. This effect is even more extreme for the weak bump case, with $w_d/H_d \sim 30$ for Z2w, due to the formation of multiple dust filaments. 
This flat morphology of dust ring is consistent with recently reported highly settled disk with relatively extended rings in Oph 163131 \citep{villenave22}, likely suggesting different turbulent levels in radial and vertical directions. 

In addition, dust feedback weakens turbulent diffusion at the dust ring center (see \S \ref{subsec:discu.interplay}).
It makes the dust ring narrower in our strong bump case, but the presence of additional off-centered filaments increases the overall $w_d$ (if they are unresolved), which largely cancels out the effect of reduced ring width.
Thus, there is a degeneracy between turbulence level, dust feedback, and bump amplitude in determining the dust ring width.
A single $\alpha$ parameter is not necessarily a good indicator of disk turbulence level, and special care is needed when interpreting the level of disk turbulence from observations of dust rings.

\paragraph{Prospects for future observations}
Overall, our simulations suggest that the dynamics of dust-trapping rings can be constrained by the presence or absence of fine-scaled substructures. Simultaneously constraining the gas bump and characterizing the fine-scale dust substructures are crucial for testify such expectations.
High resolution observations searching for kinematic features (e.g. ExoALMA \footnote{\url{https://www.exoalma.com}}) will provide essential information on the properties of the gas pressure bump, while identifying sub-$H$ scaled substructures requires observations with the longest baseline by ALMA for nearby disks, and also potentially by future upgrade of ALMA \citep[e.g.,][]{burrill22}, ngVLA \citep[e.g.,][]{andrews18b} and/or SKA \citep{ilee20}.

\section{Conclusions and Perspectives} \label{sec:conclusions}

In this paper, we conduct the first controlled experiments of dust dynamics and clumping in the MRI-turbulent pressure bumps. Our simulations include realistic level of the MRI turbulence applicable to the ambipolar-diffusion dominated outer disks, and a carefully-designed forcing scheme to achieve self-consistent balance between dust trapping and turbulent diffusion. Our main findings include:

\begin{itemize}

\item At solid abundance compatible to solar, we find robust and efficient dust clumping in radial pressure maxima, with a sizable fraction of dust mass residing in clumps over the Roche density. Secondary dust filaments may form outside of the pressure maxima, but shows only weak or no clumping.

\item Dust feedback substantially affects ring properties. Strong mass loading in dust rings reduces turbulent diffusion, making the ring narrower with stronger dust settling. Weaker pressure bump can result in formation of meta-stable secondary dust filaments near the bump.

\end{itemize}

Our work strongly support that dust-trapping rings are robust sites for planetesimal formation. Our simulations do not include self-gravity, and it is yet to examine how efficient the dust in dense clumps get converted into planetesimals. This could result from a competition between self-gravity and turbulent diffusion \citep{gerbig20}, though the internal turbulent diffusivity is uncertain (but see \cite{klahr20,klahr21}) and requires further study.

Our results also have important observational implications. In particular, depending on the strength of the gas bump, as well as dust abundance, dust feedback leads to a diversity of dust ring morphologies that may evolve over time. Conventional estimate of the level of turbulence based on ring width and dust thickness does not necessarily reflect true level of disk turbulence. Moreover, our results suggest the presence of fine-scale substructures that may be resolvable at the highest resolution by ALMA and the future ngVLA.

As a first study, we consider only a single dust species with fixed stopping time $\tau_s=0.1$. The natural next step is to incorporate a dust size distribution. This will allow us to further examine how particles of different sizes participate in dust clumping, and to yield more realistic observational signatures. Moreover, the pressure bumps in our shearing-box simulations rely on an artificial forcing prescription. Future work should consider more realistic bump formation scenarios such as magnetic flux concentration and planet-induced gas bumps in the global context.

\section*{Acknowledgements}

We thank Daniel Carrera, Kees Dullemond, Guillaume Laibe and Chao-Chin Yang for useful discussions, and Greg Herczeg for overseeing the completion of this work. This work is supported by the National Key R\&D Program of China No. 2019YFA0405100, and the China Manned Space Project with NO. CMS-CSST-2021-B09. ZX Acknowledges the support of the ERC CoG project PODCAST No. 864965 and NSFC project 11773002. Numerical simulations are conducted on the Orion cluster at Department of Astronomy, Tsinghua University, and on TianHe-1 (A) at National Supercomputer Center in Tianjin, China.

\bibliographystyle{apj}
\bibliography{ms}

\begin{thebibliography}{}
\expandafter\ifx\csname natexlab\endcsname\relax\def\natexlab#1{#1}\fi

\bibitem[{{Andrews} {et~al.}(2018{\natexlab{a}}){Andrews}, {Terrell},
  {Tripathi}, {Ansdell}, {Williams}, \& {Wilner}}]{andrews18}
{Andrews}, S.~M., {Terrell}, M., {Tripathi}, A., {et~al.} 2018{\natexlab{a}},
  \apj, 865, 157

\bibitem[{{Andrews} {et~al.}(2018{\natexlab{b}}){Andrews}, {Wilner},
  {Mac{\'\i}as}, {Carrasco-Gonz{\'a}lez}, \& {Isella}}]{andrews18b}
{Andrews}, S.~M., {Wilner}, D.~J., {Mac{\'\i}as}, E., {Carrasco-Gonz{\'a}lez},
  C., \& {Isella}, A. 2018{\natexlab{b}}, in Astronomical Society of the
  Pacific Conference Series, Vol. 517, Science with a Next Generation Very
  Large Array, ed. E.~{Murphy}, 137

\bibitem[{{Auffinger} \& {Laibe}(2018)}]{auffinger18}
{Auffinger}, J., \& {Laibe}, G. 2018, \mnras, 473, 796

\bibitem[{{Bai} \& {Stone}(2010{\natexlab{a}})}]{bai10b}
{Bai}, X.-N., \& {Stone}, J.~M. 2010{\natexlab{a}}, \apj, 722, 1437

\bibitem[{{Bai} \& {Stone}(2010{\natexlab{b}})}]{bai10c}
---. 2010{\natexlab{b}}, \apjl, 722, L220

\bibitem[{{Burrill} {et~al.}(2022){Burrill}, {Ricci}, {Harter}, {Zhang}, \&
  {Zhu}}]{burrill22}
{Burrill}, B.~P., {Ricci}, L., {Harter}, S.~K., {Zhang}, S., \& {Zhu}, Z. 2022,
  \apj, 928, 40

\bibitem[{{Carrera} {et~al.}(2015){Carrera}, {Johansen}, \&
  {Davies}}]{carrera15}
{Carrera}, D., {Johansen}, A., \& {Davies}, M.~B. 2015, \aap, 579, A43

\bibitem[{{Carrera} \& {Simon}(2022)}]{carrera22b}
{Carrera}, D., \& {Simon}, J.~B. 2022, arXiv e-prints, arXiv:2204.14270

\bibitem[{{Carrera} {et~al.}(2021){Carrera}, {Simon}, {Li}, {Kretke}, \&
  {Klahr}}]{carrera21}
{Carrera}, D., {Simon}, J.~B., {Li}, R., {Kretke}, K.~A., \& {Klahr}, H. 2021,
  \aj, 161, 96

\bibitem[{{Carrera} {et~al.}(2022){Carrera}, {Thomas}, {Simon}, {Small},
  {Kretke}, \& {Klahr}}]{carrera22a}
{Carrera}, D., {Thomas}, A.~J., {Simon}, J.~B., {et~al.} 2022, \apj, 927, 52

\bibitem[{{Chen} \& {Lin}(2020)}]{chen20}
{Chen}, K., \& {Lin}, M.-K. 2020, \apj, 891, 132

\bibitem[{{Chiang} \& {Youdin}(2010)}]{chiang10}
{Chiang}, E., \& {Youdin}, A.~N. 2010, Annual Review of Earth and Planetary
  Sciences, 38, 493

\bibitem[{{Cui} \& {Bai}(2021)}]{cui21}
{Cui}, C., \& {Bai}, X.-N. 2021, \mnras, 507, 1106

\bibitem[{{Dullemond} {et~al.}(2018){Dullemond}, {Birnstiel}, {Huang},
  {Kurtovic}, {Andrews}, {Guzm{\'a}n}, {P{\'e}rez}, {Isella}, {Zhu}, {Benisty},
  {Wilner}, {Bai}, {Carpenter}, {Zhang}, \& {Ricci}}]{dullemond18}
{Dullemond}, C.~P., {Birnstiel}, T., {Huang}, J., {et~al.} 2018, \apjl, 869,
  L46

\bibitem[{{Flaherty} {et~al.}(2018){Flaherty}, {Hughes}, {Teague}, {Simon},
  {Andrews}, \& {Wilner}}]{flaherty18}
{Flaherty}, K.~M., {Hughes}, A.~M., {Teague}, R., {et~al.} 2018, \apj, 856, 117

\bibitem[{{Flaherty} {et~al.}(2017){Flaherty}, {Hughes}, {Rose}, {Simon}, {Qi},
  {Andrews}, {K{\'o}sp{\'a}l}, {Wilner}, {Chiang}, {Armitage}, \&
  {Bai}}]{flaherty17}
{Flaherty}, K.~M., {Hughes}, A.~M., {Rose}, S.~C., {et~al.} 2017, \apj, 843,
  150

\bibitem[{{Gerbig} {et~al.}(2020){Gerbig}, {Murray-Clay}, {Klahr}, \&
  {Baehr}}]{gerbig20}
{Gerbig}, K., {Murray-Clay}, R.~A., {Klahr}, H., \& {Baehr}, H. 2020, \apj,
  895, 91

\bibitem[{{Gole} {et~al.}(2020){Gole}, {Simon}, {Li}, {Youdin}, \&
  {Armitage}}]{gole20}
{Gole}, D.~A., {Simon}, J.~B., {Li}, R., {Youdin}, A.~N., \& {Armitage}, P.~J.
  2020, \apj, 904, 132

\bibitem[{{Hashimoto} {et~al.}(2021){Hashimoto}, {Muto}, {Dong}, {Liu}, {van
  der Marel}, {Francis}, {Hasegawa}, \& {Tsukagoshi}}]{hashimoto21}
{Hashimoto}, J., {Muto}, T., {Dong}, R., {et~al.} 2021, \apj, 911, 5

\bibitem[{{Heinemann} \& {Papaloizou}(2009)}]{heinemann09}
{Heinemann}, T., \& {Papaloizou}, J.~C.~B. 2009, \mnras, 397, 64

\bibitem[{{Huang} {et~al.}(2020){Huang}, {Li}, {Isella}, {Miranda}, {Li}, \&
  {Ji}}]{huang20}
{Huang}, P., {Li}, H., {Isella}, A., {et~al.} 2020, \apj, 893, 89

\bibitem[{{Ilee} {et~al.}(2020){Ilee}, {Hall}, {Walsh}, {Jim{\'e}nez-Serra},
  {Pinte}, {Terry}, {Bourke}, \& {Hoare}}]{ilee20}
{Ilee}, J.~D., {Hall}, C., {Walsh}, C., {et~al.} 2020, \mnras, 498, 5116

\bibitem[{{Jennings} {et~al.}(2022){Jennings}, {Booth}, {Tazzari}, {Clarke}, \&
  {Rosotti}}]{jennings22}
{Jennings}, J., {Booth}, R.~A., {Tazzari}, M., {Clarke}, C.~J., \& {Rosotti},
  G.~P. 2022, \mnras, 509, 2780

\bibitem[{{Johansen} {et~al.}(2009){Johansen}, {Youdin}, \& {Mac
  Low}}]{johansen09b}
{Johansen}, A., {Youdin}, A., \& {Mac Low}, M.-M. 2009, \apjl, 704, L75

\bibitem[{{Klahr} \& {Schreiber}(2020)}]{klahr20}
{Klahr}, H., \& {Schreiber}, A. 2020, \apj, 901, 54

\bibitem[{{Klahr} \& {Schreiber}(2021)}]{klahr21}
---. 2021, \apj, 911, 9

\bibitem[{{Lehmann} \& {Lin}(2022)}]{lehmann22}
{Lehmann}, M., \& {Lin}, M.~K. 2022, \aap, 658, A156

\bibitem[{{Li} \& {Youdin}(2021)}]{li21}
{Li}, R., \& {Youdin}, A. 2021, arXiv e-prints, arXiv:2105.06042

\bibitem[{{Long} {et~al.}(2018){Long}, {Pinilla}, {Herczeg}, {Harsono},
  {Dipierro}, {Pascucci}, {Hendler}, {Tazzari}, {Ragusa}, {Salyk}, {Edwards},
  {Lodato}, {van de Plas}, {Johnstone}, {Liu}, {Boehler}, {Cabrit}, {Manara},
  {Menard}, {Mulders}, {Nisini}, {Fischer}, {Rigliaco}, {Banzatti}, {Avenhaus},
  \& {Gully-Santiago}}]{long18}
{Long}, F., {Pinilla}, P., {Herczeg}, G.~J., {et~al.} 2018, \apj, 869, 17

\bibitem[{{Manger} {et~al.}(2020){Manger}, {Klahr}, {Kley}, \&
  {Flock}}]{manger20}
{Manger}, N., {Klahr}, H., {Kley}, W., \& {Flock}, M. 2020, \mnras, 499, 1841

\bibitem[{{Nelson} {et~al.}(2013){Nelson}, {Gressel}, \& {Umurhan}}]{nelson13}
{Nelson}, R.~P., {Gressel}, O., \& {Umurhan}, O.~M. 2013, \mnras, 435, 2610

\bibitem[{{Ono} {et~al.}(2016){Ono}, {Muto}, {Takeuchi}, \& {Nomura}}]{ono16}
{Ono}, T., {Muto}, T., {Takeuchi}, T., \& {Nomura}, H. 2016, \apj, 823, 84

\bibitem[{{Paardekooper} {et~al.}(2022){Paardekooper}, {Dong}, {Duffell},
  {Fung}, {Masset}, {Ogilvie}, \& {Tanaka}}]{paardekooper22}
{Paardekooper}, S.-J., {Dong}, R., {Duffell}, P., {et~al.} 2022, arXiv
  e-prints, arXiv:2203.09595

\bibitem[{{P{\'e}rez} {et~al.}(2019){P{\'e}rez}, {Casassus}, {Baruteau},
  {Dong}, {Hales}, \& {Cieza}}]{perez19}
{P{\'e}rez}, S., {Casassus}, S., {Baruteau}, C., {et~al.} 2019, \aj, 158, 15

\bibitem[{{Pinte} {et~al.}(2016){Pinte}, {Dent}, {M{\'e}nard}, {Hales}, {Hill},
  {Cortes}, \& {de Gregorio-Monsalvo}}]{pinte16}
{Pinte}, C., {Dent}, W.~R.~F., {M{\'e}nard}, F., {et~al.} 2016, \apj, 816, 25

\bibitem[{{Raettig} {et~al.}(2021){Raettig}, {Lyra}, \& {Klahr}}]{raettig21}
{Raettig}, N., {Lyra}, W., \& {Klahr}, H. 2021, \apj, 913, 92

\bibitem[{{Riols} {et~al.}(2020){Riols}, {Lesur}, \& {Menard}}]{riols20}
{Riols}, A., {Lesur}, G., \& {Menard}, F. 2020, \aap, 639, A95

\bibitem[{{Rosotti} {et~al.}(2020){Rosotti}, {Teague}, {Dullemond}, {Booth}, \&
  {Clarke}}]{rosotti20}
{Rosotti}, G.~P., {Teague}, R., {Dullemond}, C., {Booth}, R.~A., \& {Clarke},
  C.~J. 2020, \mnras, 495, 173

\bibitem[{{Scardoni} {et~al.}(2021){Scardoni}, {Booth}, \&
  {Clarke}}]{scardoni21}
{Scardoni}, C.~E., {Booth}, R.~A., \& {Clarke}, C.~J. 2021, \mnras, 504, 1495

\bibitem[{{Sch{\"a}fer} {et~al.}(2020){Sch{\"a}fer}, {Johansen}, \&
  {Banerjee}}]{schafer20}
{Sch{\"a}fer}, U., {Johansen}, A., \& {Banerjee}, R. 2020, \aap, 635, A190

\bibitem[{{Shakura} \& {Sunyaev}(1973)}]{shakura73}
{Shakura}, N.~I., \& {Sunyaev}, R.~A. 1973, \aap, 500, 33

\bibitem[{{Stone} {et~al.}(2008){Stone}, {Gardiner}, {Teuben}, {Hawley}, \&
  {Simon}}]{stone08}
{Stone}, J.~M., {Gardiner}, T.~A., {Teuben}, P., {Hawley}, J.~F., \& {Simon},
  J.~B. 2008, \apjs, 178, 137

\bibitem[{{Suriano} {et~al.}(2018){Suriano}, {Li}, {Krasnopolsky}, \&
  {Shang}}]{suriano18}
{Suriano}, S.~S., {Li}, Z.-Y., {Krasnopolsky}, R., \& {Shang}, H. 2018, \mnras,
  477, 1239

\bibitem[{{Teague} {et~al.}(2018){Teague}, {Henning}, {Guilloteau}, {Bergin},
  {Semenov}, {Dutrey}, {Flock}, {Gorti}, \& {Birnstiel}}]{teague18}
{Teague}, R., {Henning}, T., {Guilloteau}, S., {et~al.} 2018, \apj, 864, 133

\bibitem[{{Umurhan} {et~al.}(2020){Umurhan}, {Estrada}, \& {Cuzzi}}]{umurhan20}
{Umurhan}, O.~M., {Estrada}, P.~R., \& {Cuzzi}, J.~N. 2020, \apj, 895, 4

\bibitem[{{van der Marel} {et~al.}(2021){van der Marel}, {Birnstiel}, {Garufi},
  {Ragusa}, {Christiaens}, {Price}, {Sallum}, {Muley}, {Francis}, \&
  {Dong}}]{vandermarel21}
{van der Marel}, N., {Birnstiel}, T., {Garufi}, A., {et~al.} 2021, \aj, 161, 33

\bibitem[{{Villenave} {et~al.}(2022){Villenave}, {Stapelfeldt}, {Duch{\^e}ne},
  {M{\'e}nard}, {Lambrechts}, {Sierra}, {Flores}, {Dent}, {Wolff}, {Ribas},
  {Benisty}, {Cuello}, \& {Pinte}}]{villenave22}
{Villenave}, M., {Stapelfeldt}, K.~R., {Duch{\^e}ne}, G., {et~al.} 2022, \apj,
  930, 11

\bibitem[{{Whipple}(1972)}]{whipple72}
{Whipple}, F.~L. 1972, in From Plasma to Planet, ed. A.~{Elvius}, 211

\bibitem[{{Xu} \& {Bai}(2022)}]{xu22}
{Xu}, Z., \& {Bai}, X.-N. 2022, \apj, 924, 3

\bibitem[{{Xu} {et~al.}(2017){Xu}, {Bai}, \& {Murray-Clay}}]{xu17}
{Xu}, Z., {Bai}, X.-N., \& {Murray-Clay}, R.~A. 2017, \apj, 847, 52

\bibitem[{{Yang} {et~al.}(2017){Yang}, {Johansen}, \& {Carrera}}]{yang17}
{Yang}, C.~C., {Johansen}, A., \& {Carrera}, D. 2017, \aap, 606, A80

\bibitem[{{Youdin} \& {Goodman}(2005)}]{youdin05}
{Youdin}, A.~N., \& {Goodman}, J. 2005, \apj, 620, 459

\bibitem[{{Youdin} \& {Lithwick}(2007)}]{youdin07}
{Youdin}, A.~N., \& {Lithwick}, Y. 2007, \icarus, 192, 588

\bibitem[{{Zhu} {et~al.}(2015){Zhu}, {Stone}, \& {Bai}}]{zhu15}
{Zhu}, Z., {Stone}, J.~M., \& {Bai}, X.-N. 2015, \apj, 801, 81

\end{thebibliography}

\clearpage
\begin{appendix}

\section{Dust clump properties} \label{appsec:cdf}
\setcounter{figure}{0}
\renewcommand\thefigure{\Alph{section}}

\begin{figure*}[h]
\begin{center}

\includegraphics[width=.8\textwidth]{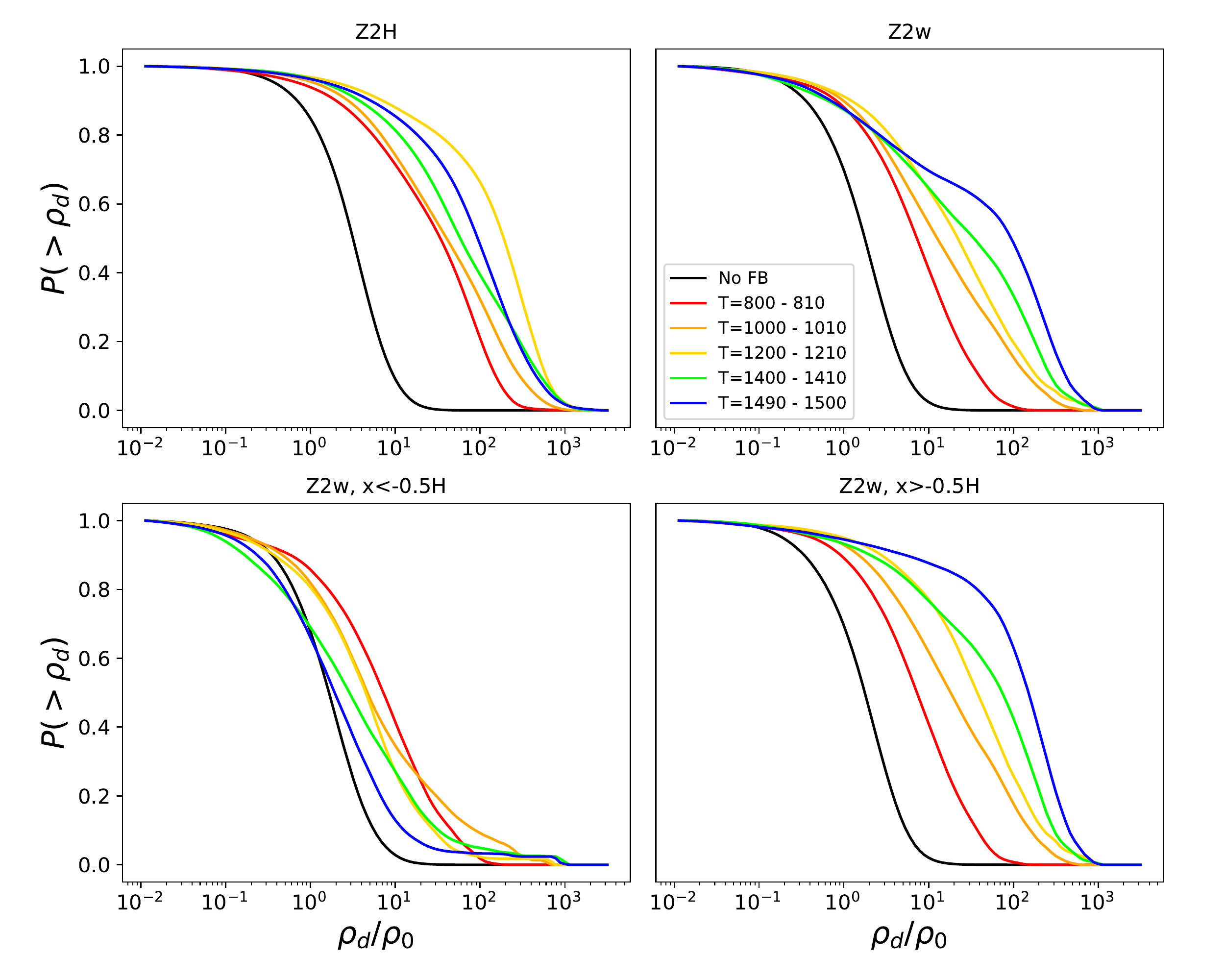}

\end{center}
\caption{\label{fig:cdf} Cumulative distribution function (CDF) of dust density for runs Z2H and Z2w, as labeled in the titles in each panel. Upper panels show CDF for the whole simulation domain, whereas the bottom two panels measure the CDFs of run Z2w for dust particles residing in $x<-0.5H$ and $x>-0.5H$ regions separately, in order to investigate clump properties in the primary and secondary rings. The CDFs are measured by averaging within different time intervals in units of $\Omega_K^{-1}$ as marked in different colors (see the legend). The black curves in each panel shows corresponding runs with no feedback, i.e., Z0 and Z0w, averaged within $T=720\Omega_K^{-1}$ to $1520 \Omega_K^{-1}$.}
\end{figure*}

In order to further analyze the dust properties in dust clumps, we present in Figure \ref{fig:cdf} the cumulative distribution function (CDF) as a function of dust density for runs Z2H and Z2w. Local dust density $\rho_d$ is evenly divided into 100 bins in logarithmic space between $0.01 \rho_0$ and 5000 $\rho_0$, and the CDF is obtained by the measuring the probability of a dust particle residing in a region with local dust density higher than the value in each bin. 

Dust density is substantially boosted for both the strong bump and weak bump cases when feedback is included, compared to the black curves showing the corresponding no feedback runs (Z0 and Z0w). The CDF is clearly time-dependent, but after about a hundred local orbits, a significant fraction of dust particles ($\gtrsim30\%$) are found to reside in regions that are $\gtrsim200$ times background gas density, reaching Roche density. Although self-gravity is not included in our simulations, such high fraction implies that planetesimal formation can be highly efficient.

For the weak bump case, we further divide the domain in half at $x=-0.5$ to separate the primary and secondary rings. We see that strong clumping occurs mostly in the primary dust ring at the bump center ($x>-0.5H$, see the lower-right panel), with the CDF very similar to the strong bump case. In the secondary ring ($x<-0.5H$, lower-left panel), the overall dust density is only boosted by a small factor $\lesssim5$ with very clumping. There are only a couple dust clumps present at later time that appear as outliers in the CDF, which represent a tiny fraction ($\lesssim3\%$) of particles. The nature of these clumps and their potential to form planetesimals deserve future studies.

\section{Dust ring evolution in long term} \label{appsec:longterm}
\setcounter{figure}{0}
\renewcommand\thefigure{\Alph{section}}

\begin{figure*}[h]
\begin{center}

\includegraphics[width=\textwidth]{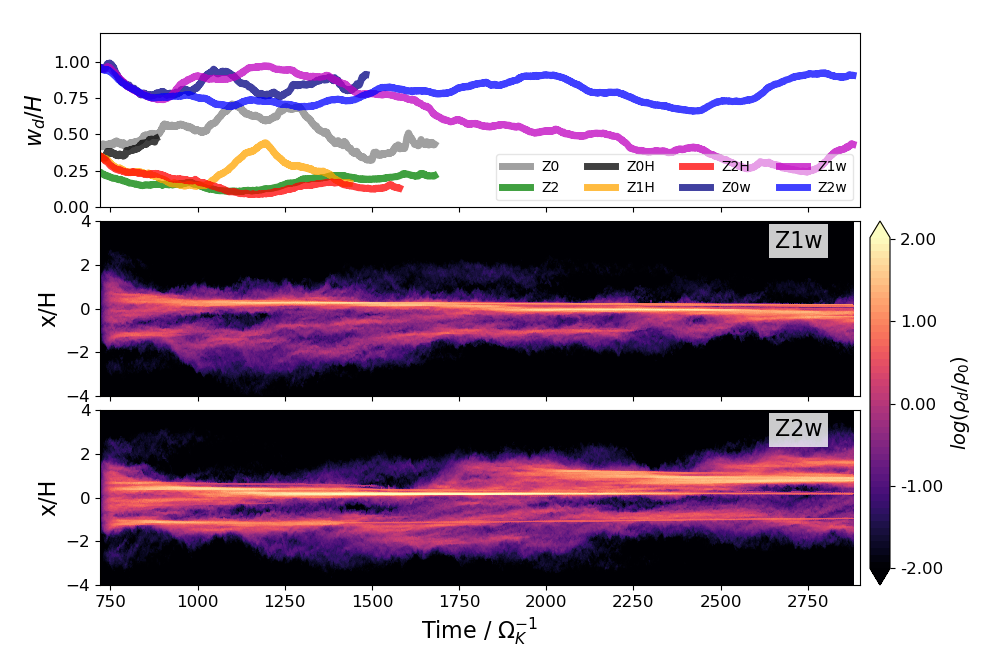}

\end{center}
\caption{\label{fig:furtherevolution} Upper panel: time evolution of dust ring width $w_d$ as a function of time $T = 720 \Omega_K^{-1}$ to $2880\Omega_K^{-1}$, featuring runs Z1w and Z2w at later stages. Lower two panels: time evolution of azimuthally and vertically averaged radial profiles of dust density at the midplane ($\pm 1$ cell from $z=0$) for runs Z1w and Z2w, same time interval as the upper panel.}
\end{figure*}

In order to examine the lifetime and long-term evolution of dust rings in our simulation, especially the secondary rings/filaments in the weak bump case, we continue to run Z1w and Z2w all the way to $T=2880 \Omega_K^{-1}$ ($T_{sim}=3000 \Omega_K^{-1}$), and show the evolution of dust ring widths and dust ring profiles in Figure \ref{fig:furtherevolution}.

For run Z1w, shown in the upper panel of Figure \ref{fig:furtherevolution}, the overall ring width decreases after $T\sim1500\Omega_K^{-1}$.
This can also be seen clearly in the lower panel of Figure \ref{fig:furtherevolution}, showing the evolution of dust radial distribution, where the secondary ring spreads and later merge into the primary ring. On the other hand, the secondary ring in run Z2w is more long-lived, persisting over hundreds of orbits. This is likely due to the higher inertia in the dust ring which withstands potential disruption by turbulent diffusion (see also Appendix \ref{appsec:diff}) and/or radial drift. It is also in line with our results in \citetalias{xu22}, where dust feedback enhances the MRI zonal flows with more extended lifetimes. Moreover, we see that additional secondary rings/filaments may form at off-center positions, and even the primary ring itself made a small shift in its position over the duration of the long-term simulation (while still residing in the bump center). The results indicate that the presence of secondary rings/filaments is likely generic, which dynamically evolve over the timescale of tens to a few hundreds orbits.

\clearpage
\section{Turbulent diffusion in dust rings} \label{appsec:diff}
\setcounter{figure}{0}
\renewcommand\thefigure{\Alph{section}.\arabic{figure}}

\begin{figure*}[h]
\begin{center}
\includegraphics[width=0.47\textwidth]{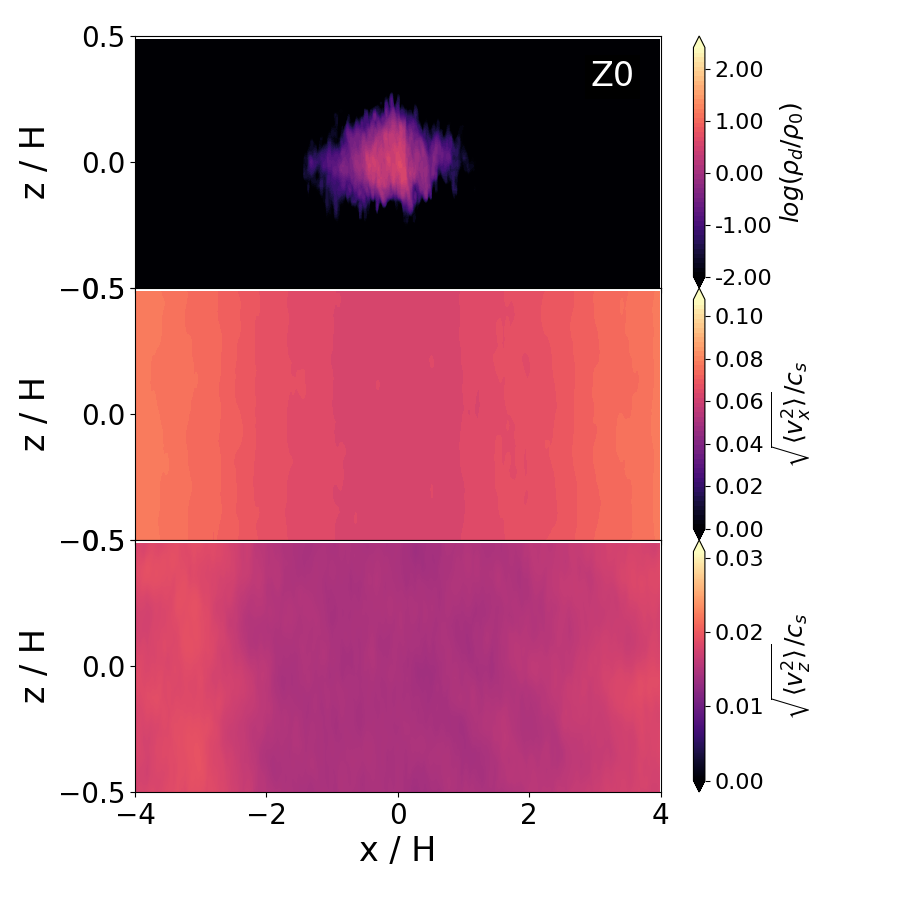}
\includegraphics[width=0.47\textwidth]{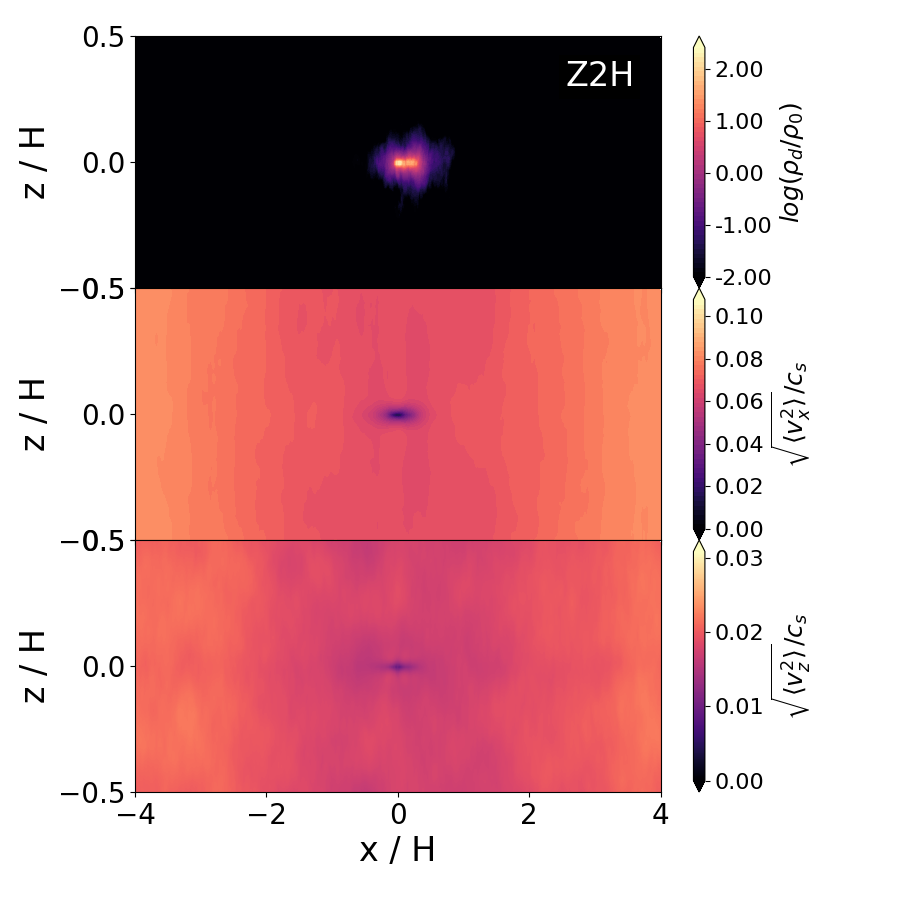}
\end{center}

\caption{Azimuthally averaged spatial distribution of dust density (upper panels in each plot), gas rms velocity on radial direction (middles panels) and vertical direction (lower panels), time-averaged within $T=800 - 1500 \Omega_K^{-1}$, for runs Z0 (left panels) and Z2H (right panels). Note that this figure is not to the real aspect ratio.\label{fig:diff}}
\end{figure*}

\begin{figure*}[h]
\begin{center}
\includegraphics[width=0.47\textwidth]{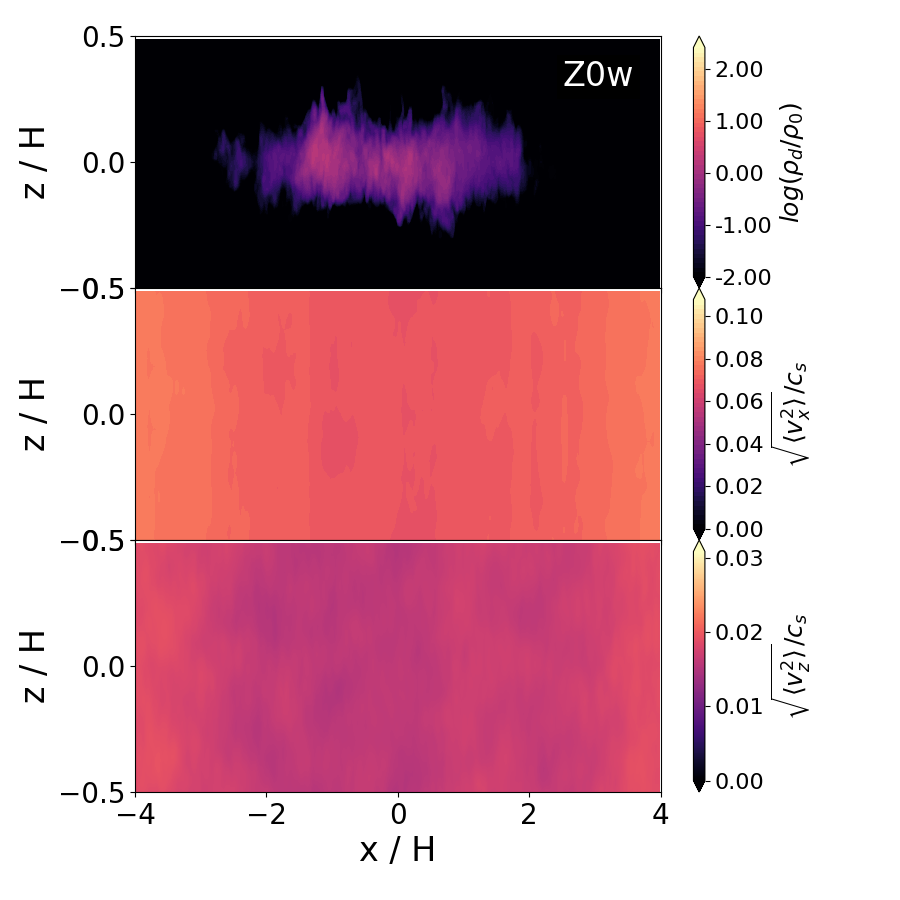}
\includegraphics[width=0.47\textwidth]{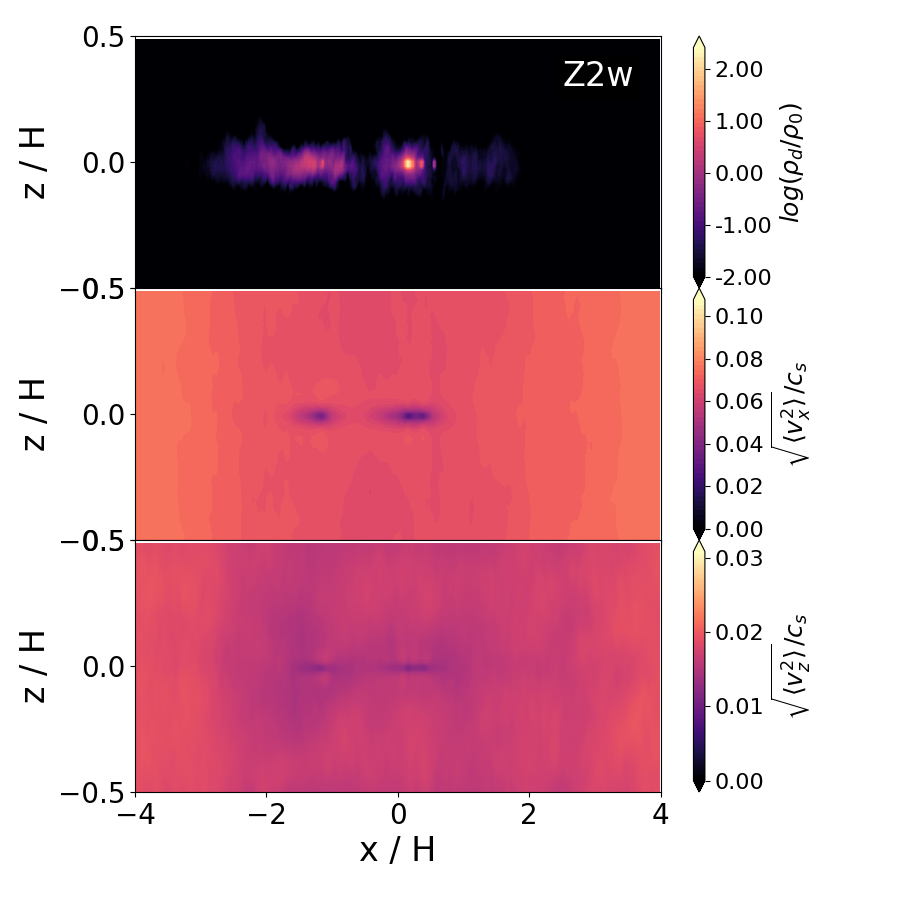}
\end{center}

\caption{Same as Figure \ref{fig:diff}, but for runs Z0w and Z2w. \label{fig:diffw}}
\end{figure*}

Figure \ref{fig:diff} and Figure \ref{fig:diffw} show the azimuthal and time averaged gas rms velocities in the radial and vertical directions, i.e. $\sqrt{\langle v_x^2 \rangle}$ and $\sqrt{\langle v_z^2 \rangle}$, for the strong and weak bump cases respectively. They serve as a proxy for estimating the strength of turbulent diffusion.
For run Z0 without dust feedback, gas rms velocities in both directions are approximately uniform, with $\sqrt{\langle v_x^2 \rangle} \sim 0.07 c_s$ and $\sqrt{\langle v_z^2 \rangle} \sim 0.016 c_s$ (the results from run Z0w are largely the same). 
With $w_d/w_{\rm bump} = 0.28$ and $H_d/H = 0.067$, our results are overall consistent with the expectations of $w_d/w_{\rm bump}\sim\sqrt{\langle v_x^2 \rangle/(\Omega_K \tau_s)}$ and $H_d/H \sim \sqrt{\langle v_z^2 \rangle/(\Omega_K \tau_s)}$, modulo correction factors from the eddy time \citep[][\citetalias{xu22}]{youdin07,zhu15}. They are also generally consistent with the turbulence parameter $\alpha \sim10^{-3}$ measured in our simulation.\footnote{The MRI turbulence level in our simulations can be characterized by $\alpha$ parameter \citep{shakura73}, and is obtained by summing over the Maxwell and Reynold stresses normalized by thermal pressure. Our simulation without dust feedback (Z0) gives time-averaged $\alpha = 1.08 \times 10^{-3}$ , consistent with previous studies \citep[][\citetalias{xu22}]{zhu15,xu17}.}

With feedback included in run Z2H, we find that the radial gas rms velocity is largely reduced in the dust ring, which is clearly visible in Figure \ref{fig:diff}. We obtain $\sqrt{\langle v_x^2 \rangle} \sim 0.07 c_s$ by averaging within the whole domain, which is consistent to the value measured in Z0. In the the dust ring, however, we find $\sqrt{\langle v_{x,{\rm ring}}^2 \rangle} \sim 0.02 c_s$, about 3 times lower than that outside of the ring. This reduction in $\sqrt{\langle v_x^2 \rangle}$ generally matches the reduction in ring width reported earlier.
In the vertical direction, we note that in \citetalias{xu22}, we found that dust feedback leads to a strong reduction in the turbulent eddy time which promotes settling. In this work, we find that in addition, with more enhanced dust mass loading, the vertical rms velocity $\sqrt{\langle v_z^2 \rangle}$ is also reduced. Although the reduction is less dramatic than that in the radial direction, it is sufficient to account for the additional settling found in Figure \ref{fig:hist}. The reduction in both $\sqrt{\langle v_x^2 \rangle}$ and $\sqrt{\langle v_z^2 \rangle}$ is also seen in run Z2w (see right panels of Figure \ref{fig:diffw}).

\end{appendix}

\end{document}